\RequirePackage{fix-cm}
\documentclass[twocolumn,epjc3]{svjour3}  
\RequirePackage{graphicx}
\RequirePackage{amsmath,amssymb}
\RequirePackage{mathptmx}
\RequirePackage{isotope}
\RequirePackage{cite}
\usepackage{upgreek}
\usepackage{color}
\usepackage{doi}
\RequirePackage{hyperref}
\RequirePackage{booktabs}
\RequirePackage{siunitx}
\RequirePackage{lineno}\newdimen\linenumbersep\linenumbersep=1.5pt\renewcommand{\linenumberfont}\tiny
\journalname{Eur. Phys. J. C}
\newcommand{\kr}{\mbox{$^{83\mathrm{m}}$Kr}}
\begin{document}
\title{Calibration of high voltages at the ppm level by the difference of \kr\ conversion electron lines at the KATRIN experiment}
%
%
%

\institute{%
Helmholtz-Institut f\"{u}r Strahlen- und Kernphysik, Rheinische Friedrich-Wilhelms Universit\"{a}t Bonn, Nussallee 14-16, 53115 Bonn, Germany\label{a}
\and Institute of Experimental Particle Physics~(ETP), Karlsruhe Institute of Technology~(KIT), Wolfgang-Gaede-Str. 1, 76131 Karlsruhe, Germany\label{b}
\and Institut f\"{u}r Physik, Johannes Gutenberg-Universit\"{a}t Mainz, 55099 Mainz, Germany\label{c}
\and Institute for Data Processing and Electronics~(IPE), Karlsruhe Institute of Technology~(KIT), Hermann-von-Helmholtz-Platz 1, 76344 Eggenstein-Leopoldshafen, Germany\label{d}
\and Institute for Nuclear Research of Russian Academy of Sciences, 60th October Anniversary Prospect 7a, 117312 Moscow, Russia\label{e}
\and Institute for Technical Physics~(ITeP), Karlsruhe Institute of Technology~(KIT), Hermann-von-Helmholtz-Platz 1, 76344 Eggenstein-Leopoldshafen, Germany\label{f}
\and Max-Planck-Institut f\"{u}r Kernphysik, Saupfercheckweg 1, 69117 Heidelberg, Germany\label{g}
\and Max-Planck-Institut f\"{u}r Physik, F\"{o}hringer Ring 6, 80805 M\"{u}nchen, Germany\label{h}
\and Technische Universit\"{a}t M\"{u}nchen, James-Franck-Str. 1, 85748 Garching, Germany\label{i}
\and Institute for Nuclear Physics~(IKP), Karlsruhe Institute of Technology~(KIT), Hermann-von-Helmholtz-Platz 1, 76344 Eggenstein-Leopoldshafen, Germany\label{j}
\and Laboratory for Nuclear Science, Massachusetts Institute of Technology, 77 Massachusetts Ave, Cambridge, MA 02139, USA\label{k}
\and Center for Experimental Nuclear Physics and Astrophysics, and Dept.~of Physics, University of Washington, Seattle, WA 98195, USA\label{l}
\and Nuclear Physics Institute of the CAS, v. v. i., CZ-250 68 \v{R}e\v{z}, Czech Republic\label{m}
\and Institut f\"{u}r Kernphysik, Westf\"{a}lische Wilhelms-Universit\"{a}t M\"{u}nster, Wilhelm-Klemm-Str. 9, 48149 M\"{u}nster, Germany\label{n}
\and Department of Physics, Faculty of Mathematics and Natural Sciences, University of Wuppertal, Gauss-Str. 20, D-42119 Wuppertal, Germany\label{o}
\and Department of Physics, Carnegie Mellon University, Pittsburgh, PA 15213, USA\label{p}
\and Universidad Complutense de Madrid, Instituto Pluridisciplinar, Paseo Juan XXIII, n\textsuperscript{\b{o}} 1, 28040 - Madrid, Spain\label{q}
\and Department of Physics and Astronomy, University of North Carolina, Chapel Hill, NC 27599, USA\label{r}
\and Triangle Universities Nuclear Laboratory, Durham, NC 27708, USA\label{s}
\and Commissariat \`{a} l'Energie Atomique et aux Energies Alternatives, Centre de Saclay, DRF/IRFU, 91191 Gif-sur-Yvette, France\label{t}
\and University of Applied Sciences~(HFD)~Fulda, Leipziger Str.~123, 36037 Fulda, Germany\label{u}
\and Department of Physics, Case Western Reserve University, Cleveland, OH 44106, USA\label{v}
\and Institute for Nuclear and Particle Astrophysics and Nuclear Science Division, Lawrence Berkeley National Laboratory, Berkeley, CA 94720, USA\label{w}
\and Institut f\"{u}r Physik, Humboldt-Universit\"{a}t zu Berlin, Newtonstr. 15, 12489 Berlin, Germany\label{x}
\and Project, Process, and Quality Management~(PPQ), Karlsruhe Institute of Technology~(KIT), Hermann-von-Helmholtz-Platz 1, 76344 Eggenstein-Leopoldshafen, Germany    \label{y}
}

\author{%
M.~Arenz\thanksref{a}
\and W.-J.~Baek\thanksref{b}
\and M.~Beck\thanksref{c}
\and A.~Beglarian\thanksref{d}
\and J.~Behrens\thanksref{b}
\and T.~Bergmann\thanksref{d}
\and A.~Berlev\thanksref{e}
\and U.~Besserer\thanksref{f}
\and K.~Blaum\thanksref{g}
\and T.~Bode\thanksref{h,i}
% ...
}

\thankstext{rest}{e-mail: oliver.rest@uni-muenster.de}
\thankstext{also1}{Also affiliated with Oak Ridge National Laboratory, Oak Ridge, TN 37831, USA}

\date{Received: 20 February 2018 / Accepted: 22 April 2018}
\maketitle
\newpage
\begin{abstract}
The neutrino mass experiment KATRIN requires a stability of 3~ppm for the retarding potential at -18.6~kV of the main spectrometer. To monitor the stability, two custom-made ultra-precise high-voltage dividers were developed and built in cooperation with the German national metrology institute Physikalisch-Technische Bundesanstalt (PTB). Until now, regular absolute calibration of the voltage dividers required bringing the equipment to the specialised metrology laboratory. Here we present a new method based on measuring the energy difference of two \kr\ conversion electron lines with the KATRIN setup, which was demonstrated during KATRIN's commissioning measurements in July 2017. The measured scale factor $M=1972.449(10)$ of the high-voltage divider K35 is in agreement with the last PTB calibration four years ago. This result demonstrates the utility of the calibration method, as well as the long-term stability of the voltage divider.
\end{abstract}
\section{Introduction}
Precision high voltages (HV) at the ppm level are required for many applications in science, e.g. for defining the kinetic energy of electrons in an electron cooler at storage rings~\cite{ullmann} or for the precise determination of the energy of electrons in electrostatic retarding spectrometers or other analysers~\cite{Dragoun2004,xpsbook,PICARD1992345}.

\begin{figure*}[ht]
\center
\includegraphics[width=1.0\textwidth]{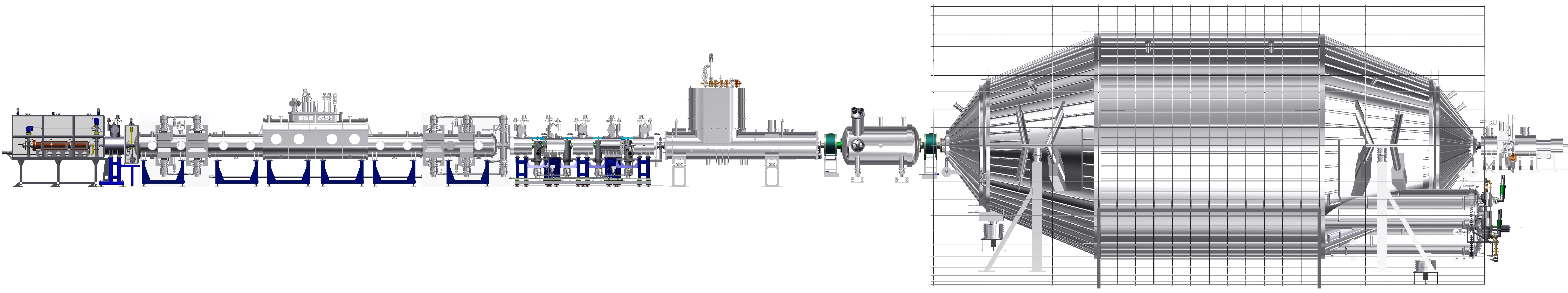}
\caption{Experimental setup of the KATRIN experiment. The main components are (from left to right): Calibration and monitoring rear section, windowless gaseous tritium source, transport section, pre-spectrometer, main spectrometer and focal plane detector.}
\label{KATRIN_beamline}
\end{figure*}

The KArlsruhe TRitium Neutrino (KATRIN) experiment~\cite{designreport} at the Karlsruhe Institute of Technology (KIT) (see fig.\ref{KATRIN_beamline}) aims for a direct neutrino mass determination by a precise measurement of the tritium-$\upbeta$-decay spectrum near the endpoint. The expected sensitivity of the experiment is 0.2~eV/$c^2$ at 90\%~C.L.~\cite{Drexlin:2013lha}. Currently the Mainz-~\cite{Kraus:2004zw} and Troitsk-~\cite{Lobashev:2003kt,Aseev:2011dq} neutrino mass experiments set upper limits on the neutrino mass of 2~eV/c$^2$.\\
In KATRIN, electrons are emitted from molecular tritium decaying in the windowless gaseous tritium source (WGTS) and are guided adiabatically by magnetic fields through the transport section. In this transport section, tritium is removed from the beamline by means of differential and cryogenic pumping. 
In the pre- and main spectrometers downstream from the transport section, the kinetic energy of the electrons is analysed. 
In order to reach the desired sensitivity, the spectrometers need to provide a large acceptance angle for the emitted $\upbeta$-electrons and, in the case of the main spectrometer, a very good energy resolution as well. 
This is accomplished by operating the spectrometers as MAC-E filters~\cite{PICARD1992345}, which are electrostatic retardation spectrometers (high energy filters) combined with magnetic adiabatic collimation to obtain large solid angle acceptance.  
Following the spectrometers, the focal-plane detector (FPD)~\cite{AMSBAUGH201540} counts all electrons that have sufficient energy to pass both spectrometers. This FPD is a monolithic silicon $p$-$i$-$n$ diode with 148 pixels.

One key requirement of the experiment is the stability of the retarding potential ($U_{\text{ret}}\approx$-18.6~kV) of the main spectrometer, which has to be maintained and monitored with a precision of 3~ppm (60~mV) over the measurement periods of two months. The knowledge of the absolute retarding potential would additionally allow a comparison of the tritium-$\upbeta$ endpoint with the nuclear mass difference of $^3$He and tritium $^3$H determined with Penning traps~\cite{Streubel2014,VanDyck:1993zz,Myers:2015lca}.  

Two independent approaches to monitor the high voltage are being pursued in order to ensure system redundancy.  Firstly, since the HV cannot be measured directly with the required precision, a voltage divider is used to scale the retarding potential to $\lesssim 20$~V. The scaled retarding potential can then be determined with a commercial precision digital voltmeter (DVM). In this range, the voltage measurement can be calibrated against a 10~V reference, based on the Josephson effect, at the German national metrology institute Physikalisch-Technische Bundesanstalt (PTB). 
Since there is no commercial solution available with the required precision and stability, two custom-made ultra-precise high voltage dividers, named K35~\cite{Thummler:2009rz} and K65~\cite{Bauer:2013pca}, were developed and built in cooperation with the PTB. 

An HV divider is characterised by its scale factor $M$, which is defined as the ratio of input and output voltages. An absolute calibration with ppm-precision can usually be performed exclusively at metrology centres.  The scale factor of the K35 was measured at PTB in 2013 to be
\begin{equation}
\label{PTB_value}
M_{\text K35}^{\text{PTB,\,2013}} = 1972.4531(20)\, .
\end{equation}

Secondly, the HV is compared to a natural standard given by mono-energetic conversion electrons from the decay of \kr. The parent isomer $^{83}$Rb is implanted into a solid-state source~\cite{1748-0221-8-03-P03009} at the BONIS facility in Bonn \cite{arenz_diss} and used at the monitor spectrometer~\cite{Erhard:2014ema}. This third MAC-E filter is connected to the HV of the main spectrometer.
Due to solid-state and surface effects, such as electron energy losses in the source material and drifts of the work function, absolute calibration of the HV at the required precision is not possible with such a source. However, relative changes over a measurement period of up to several months can be monitored~\cite{slezak_diss}.

In July 2017, a calibration and measurement campaign with gaseous \kr, injected into the WGTS from a \kr\ generating $^{83}$Rb source~\cite{1742-6596-888-1-012072}, was performed with the complete KATRIN beamline~\cite{Drexlin:2013lha}. With the well-known energies of mono-energetic conversion electron lines of this isotope, source properties of the WGTS and transmission properties of the spectrometers were investigated. Furthermore, the adiabatic transport of electrons from the source to the detector and the general alignment and functionalities of the complete system were tested~\cite{kr_paper, kr_paper_2}.  This measurement campaign also provided the opportunity to calibrate the K35 HV divider to the ppm-level by comparing two conversion electron lines.
A similar HV calibration was previously performed using a condensed \kr\ source at the former Mainz neutrino mass experiment~\cite{thummler_diss}. The main idea is to compare the kinetic energy of conversion electrons emitted from the same nuclear transition, but originating from different atomic shells. The systematic uncertainty of the nuclear transition energy cancels; the only remaining uncertainty, which is an order of magnitude lower, arises from the atomic binding energies. The measurements reported in~\cite{thummler_diss} were limited by systematic corrections of the order 100~meV.  These corrections, which are not precisely known, account for the final state effects of the decaying nucleus in a submonolayer of \kr\ on a highly oriented pyrolytic graphite (HOPG) substrate. Gaseous sources overcome the disadvantages of a condensed or solid-state source.  
 
In this work, the calibration of the HV divider K35 with gaseous \kr\ is presented.  The next section gives an overview of the calibration concept and the determination of the scale factor of the divider using \kr\ conversion electron line measurements. Subsequently, the results of the calibration measurements performed at KATRIN are reported.
\section{Calibration of a HV divider with \kr\ conversion electrons}
\label{section_2}
\kr\ decays via two cascaded transitions with gamma energies of 32151.6(5)~eV and 9405.7(6)~eV, respectively~\cite{MCCUTCHAN2015201}. Both transitions decay dominantly by emission of conversion electrons instead of gamma radiation. In this work, only conversion electrons from the 32~keV transition are used. 

The kinetic energy $E_{\text{kin}}$ of a conversion electron from a \kr\ atom decaying freely in vacuum depends on the energy of the transition $E_{\upgamma}$, the atomic binding energy $E_{\text{bin}}$, the nuclear recoil energies caused by the gamma\footnote{The nuclear recoil energy of the gamma transition $E_{\text{rec}}^{\upgamma}$ enters, since the nuclear transition energy $\Delta E_\mathrm{fi}$ and the tabulated gamma energy $E_\upgamma$ differ by this nuclear recoil energy:  $\Delta E_\mathrm{fi} - E_{\text{rec}}^{\upgamma} = E_\upgamma$.} $E_{\text{rec}}^{\upgamma}$, 
and the conversion electron $E_{\text{rec}}^{\text{ce}}$:
\begin{equation}
\label{kin_en}
E_{\text{kin}} = E_{\upgamma} - E_{\text{bin}} + E_{\text{rec}}^{\upgamma} - E_{\text{rec}}^{\text{ce}}.
\end{equation}
The binding energy depends on the atomic shell of the emitted electron.
The values for $E_{\text{bin}}$ used in this analysis were determined with X-ray and photoelectron spectroscopy measurements~\cite{Dragoun2004} to be 14327.26(4)~eV for the K- and 1679.21(3)~eV for the L$_3$-subshell.
The nuclear recoil energy of the conversion electrons $E_{\text{rec}}^{\text{ce}}$ can be calculated to be 0.120~eV for the K- and 0.207~eV for the L$_3$-subshell for the 32~keV transition, both with negligible uncertainty.

\kr\ decays in the WGTS under ultra-high vacuum conditions. The conversion electrons are guided magnetically and adiabatically through the beamline to the main spectrometer, where an integrated spectrum is recorded by varying the retarding potential. With the HV divider K35 and a precision digital voltmeter (Fluke~8508A)\footnote{The DVM was calibrated with a PTB-calibrated 10~V reference device (Fluke~732A).}, which measures the output voltage $U_{\text{DVM}}$ of the K35, the corresponding retarding energy $qU$ for a particle of charge $q$  can be determined using
\begin{equation}
\label{u_ret_m}
q \cdot U = q \cdot U_{\text{DVM}} \cdot  M_{\text{K35}}, 
\end{equation}
where $q=-$e.  The transmission condition for electrons to pass the main spectrometer is given by
\begin{equation}
\label{trans_con}
E_{\text{kin}} \geq q \cdot U_{\text{DVM}} \cdot  M_{\text{K35}} - \Delta \Phi - q \cdot  U_{\text{pot.dec.}}.
\end{equation}
Since the retarding voltage is applied between the Fermi energies of the source tube and the spectrometer electrode system, $\Delta \Phi$ describes the correction for the difference between the work functions of the two materials. Due to the large diameter of the main spectrometer of 10~m and the wire electrode covering the inner surface \cite{Valerius:2010zz}, the retarding potential across the analysing plane is not perfectly equal to the applied potential and shows a radial dependence. Therefore, a pixel-wise correction for the potential $U_{\text{pot.dec.}}$ has been used. This correction amounts to about 2.25~V for the 40 innermost detector pixels with a r.m.s. value of less than 60~mV and scales nearly linearly in radial direction. The average difference of this correction over all these pixels amounts to 9~mV between the HV settings for the K-32 and the L$_3$-32 measurement. 

The so-called transmission edge is a special case where the kinetic energy of the electrons equals the right-hand side of equation \ref{trans_con}. Using equations \ref{kin_en} and \ref{trans_con}, the scale factor of the HV divider is then given as
\begin{equation}
\label{M_single}
M_{\text{K35}} = \frac{E_{\upgamma} - E_{\text{bin}} + E_{\text{rec}}^{\upgamma} - E_{\text{rec}}^{\text{ce}} + \Delta \Phi + q \cdot U_{\text{pot.dec.}}}{q \cdot U_{\text{DVM}}}
\end{equation}
with $U_{\text{DVM}}$ measured at the transmission edge.
Following equation \ref{M_single}, the K35 could be calibrated by analysing just a single line position. However, the nuclear transition energy and the work function difference are not known to the desired ppm level. This limitation can be resolved using the energy difference of two conversion electron lines from the same gamma transition
\begin{equation}
\label{M_multi}
M_{\text{K35}} = \frac{\Delta E_{\text{bin}} + \Delta E_{\text{rec}}^{\text{ce}} + q \cdot \Delta U_{\text{pot.dec.}}}{q \cdot \Delta U_{\text{DVM}}}
\end{equation}
so that $E_\upgamma$ and $\Delta\Phi$\footnote{The work functions of the source and spectrometer should be constant on the time scale of the measurements.} are eliminated from the equation\footnote{With a very different technique \cite{noertershaeuser2018}, collinear laser spectroscopy on Doppler-shifted ions, an absolute high voltage calibration has been reported recently. This novel method profits from eliminating systematics by performing two measurements at two different Doppler-shifts similarly to the method reported here.}.
Since K35 has a negligible voltage dependency of 0.03~ppm/kV \cite{Thummler:2009rz}, we assume a constant scale factor for the HV settings of the K-32 and the L$_3$-32 measurement.
The differences of the binding and recoil energies
\begin{align}
&\Delta E_{\text{bin}} = E_{\text{bin}}^{\text{L}_3} - E_{\text{bin}}^{\text{K}} \text{,} \\
&\Delta E_{\text{rec}}^{\text{ce}} = E_{\text{rec}}^{\text{ce, L}_3\text{-}32} - E_{\text{rec}}^{\text{ce, K-32}} 
\end{align}
add up to 
\begin{equation}
\label{eq:delta_ebin}
\Delta E_{\text{bin}} + \Delta E_{\text{rec}}^{\text{ce}}  = 12647.963(50)_\mathrm{sys}~\mathrm{eV}\, . 
\end{equation}
The potential correction $U_{\text{pot.dec.}}$ has been determined for every FPD pixel by an electric field calculation with the simulation software Kassiopeia \cite{1367-2630-19-5-053012}.

In order to determine the individual line energy positions, the observed integral spectrum was fitted with MINUIT \cite{refMINUIT}. The fit function consists of a Lorentzian with free amplitude $a$, width $\Gamma$ and the energy $E(\text{K-32})$ or $E(\text{L}_3\text{-}32)$, convolved with the transmission function $T(E, U_{\text{DVM}})$ of the main spectrometer:
\begin{align}
\label{trans_fun}
&T(E, U_{\text{DVM}})= \nonumber \\
&\left\{
\begin{array}{ll}
0 & \text{ for } E - q U \leq 0\\
\frac{1 - \sqrt{1 - \frac{E - q U}{E} \frac{B_{\text{s}}}{B_{\text{a}}}  \cdot \frac{2}{\gamma + 1}}}{1 - \sqrt{1 - \frac{B_{\text{s}}}{B_{\text{m}}}}} & 
	 \text{ for }  0 < E - q U < \Delta E_\mathrm{trans}\\
1 &  \text{ for }  E - q U \geq \Delta E_\mathrm{trans}
\end{array}  
\right
.
\end{align}
and a constant background term $b$~\cite{Drexlin:2013lha}. Here we used the abbreviation $U = U_{\text{DVM}} \cdot  M_{\text{K35}}$ from equation \ref{u_ret_m}. Relativistic corrections are included in equation~\ref{trans_fun} using the Lorentz factor $\gamma$ of the electron. 
The width of the transmission function $\Delta E_\mathrm{trans} = E \cdot \frac{B_{\text{a}}}{B_{\text{m}}} \cdot \frac{\gamma + 1}{2}$ is calculated from the energy of the electrons and the ratio of the magnetic flux densities in the analysing plane ($B_{\text{a}} = 0.268$~mT) and that at the exit of the spectrometer ($B_{\text{m}}=4.20$~T). 

Equation~\ref{trans_fun} is derived assuming an isotropically emitting source. In the case that the source magnetic field ($B_{\rm s}=2.52$~T) is lower than the maximum magnetic field encountered by the electrons on their way to the detector, the maximum amplitude of the transmission function is limited by magnetic reflection of the electrons with emission angles that exceed a cut-off angle given by 
\begin{equation}
\theta^{\rm max}_{\rm start} = \arcsin\left(\sqrt{\frac{B_{\rm s}}{B_{\rm m}}}\right) \approx 50.8^\circ.
\end{equation}
The final fit function for the K-32 line is
\begin{align}
&f(E)= \nonumber \\
&\int\limits_{q U_{\text{DVM}} M_{\text{K35}}}^{\infty} \frac{a / \pi \cdot \Gamma/2}{(E(\text{K-32})-E^{\prime})^2+\Gamma^2/4} \cdot T^{\prime}(E^{\prime} ,U_{\text{DVM}}) \ \mathrm{d}E^{\prime} + b \, ,
\end{align}
where the modified transmission function $T^{\prime}(E^{\prime} ,U_{\text{DVM}})$ contains three additional corrections: Firstly, the temperature of the \kr\ gas in the WGTS of 100~K leads to a thermal Gaussian broadening of the conversion line. Secondly, a high voltage ripple of the retarding potential was observed throughout the measurements\cite{kr_paper}.  The ripple had a nearly sinusoidal shape with a frequency of 50~Hz and an amplitude of 187~mV for the K-32 line and of 208~mV for the L$_3$-32 line. The Gaussian broadening and the recorded ripple signal are convolved with the transmission function in the fit.
Thirdly, the shape of the transmission function will be modified by synchrotron radiation losses\footnote{The synchrotron radiation affects the transversal energy $E_\perp$ (i.e. in the motion direction transversal of the magnetic field $B$) with the power loss 
$\dot E_\perp = - \frac{e^4 \beta^2 \gamma^2}{6 \pi \varepsilon_0 m^2_e c} \cdot B^2$, where $\gamma$ is the relativistic factor and $\beta = v_\perp/c$ is the velocity.  In the non-relativistic case, the power loss amounts to $\dot E_\perp = - \frac{0.39}{[\mathrm{T^2 s}]} \cdot E_\perp \cdot B^2$. Hence electrons emitted with high angles will be transmitted at lower retarding potentials, which results in a broadening of the transmission function.}, increasing its width $\Delta E_\mathrm{trans}$  by about 3~\% (2~\%) for the K-32 (L$_3$-32) line.
\section{Calibration results for the HV divider K35}
During the KATRIN calibration and measurement phase in July 2017, the energy of all conversion electron lines of the gaseous \kr\ source were measured. The K-32 and L$_3$-32 lines were used to calibrate the high-voltage divider K35 as described in section~\ref{section_2}.

\begin{figure*}[ht]
\center
\includegraphics[width=1.0\textwidth]{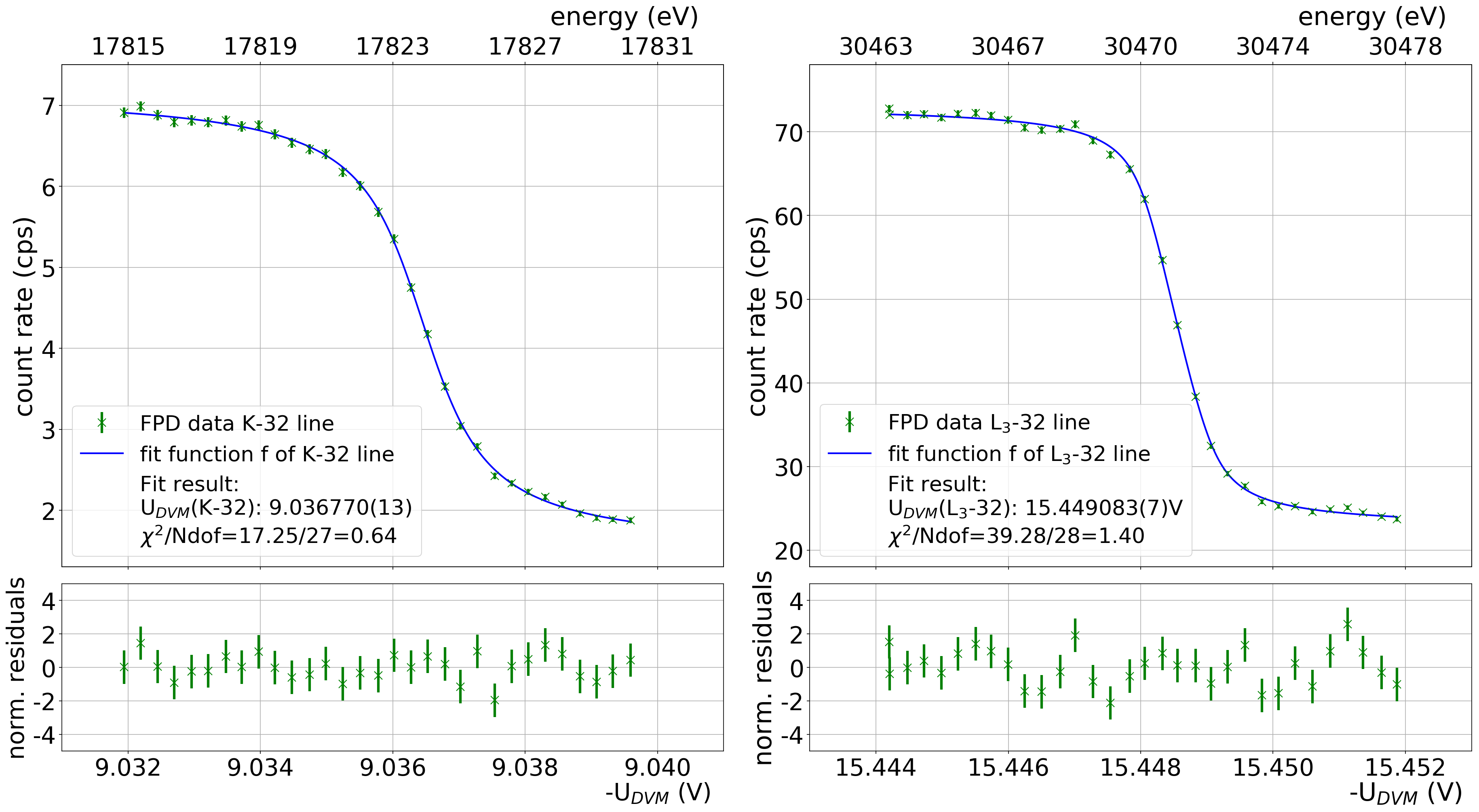}
\caption{Averaged K-32 (left) and L$_3$-32 (right) data of the innermost 40 detector pixels with a four-parameter line fit to visualize the more complex constrained fit of the 40 individual pixels (see text). The denoted line position does not include the work function difference $\Delta \Phi$ between the source and the spectrometer, which is not known to ppm precision, or other systematic uncertainties.
The upper abscissa provides the corresponding energies according to equations \ref{PTB_value} and \ref{u_ret_m}.
The panels below the fits display the normalised residuals.}
\label{bild_fit}
\end{figure*}

In this work, a combined analysis of the 40 innermost detector pixels (out of 148 pixels in total) was performed to obtain high statistics while avoiding increased systematic uncertainties at larger beam radii. Each detector pixel was treated with its corresponding potential correction $U_{\text{pot.dec.}}$. For illustration, the average of all K-32 and L$_3$-32 conversion electron data of these 40 innermost detector pixels has been calculated and fitted, as shown in figure \ref{bild_fit}.
The good agreement between data and the fit model can be seen in the residuals as well as in the reduced $\chi ^2$ values of the fits.

For the final result, we avoid averaging the pixel-dependent $U_{\text{pot.dec.}}$ values by performing a combined 82-parameter fit\footnote{In this combined fit we used common fit parameters for line width $\Gamma$ and position $E(\text{K-32})$ (or $E(\text{L}_3\text{-}32)$, respectively) but with separate fit parameters for amplitude $a_{i}$ and background $b_i$ for each pixel ($i=1,...,40$).} 
of the data from the 40 innermost detector pixels, leading to the results shown in table \ref{table_results}.
\begin{table}[!!!!h]
\center
\caption{Fit result for the combined analysis of the 40 innermost detector pixels. For the denoted line, the work function difference $\Delta \Phi$ between the source and the spectrometer is not known with sufficient precision and is not included. The same holds for other systematic uncertainties. We estimate the uncertainty of $\Delta \Phi$ to be of the order of a few 100~meV. As $\Delta \Phi$ drops out in the calculation of $\Delta U_{\text{DVM}}$, this does not pose a problem for further analyses.
\label{table_results}}
\begin{tabular}{lll}
\toprule
parameter & K-32 & L$_3$-32 \\ 
\toprule
U$_{\text{DVM}}$ line position (V) & 9.036768(12) & 15.449083(9)\\ 
\midrule
U$_{\text{DVM}}$ line width (V) & 0.00135(4) & 0.00056(2) \\ 
\midrule
$\chi^2$/N$_{\text{dof}}$ & $\frac{1131.91}{1158}$=0.98 & $\frac{1257.51}{1198}$=1.05 \\ 
\bottomrule
\end{tabular}
\end{table}\\
The results from table~\ref{table_results} yield a voltage difference of
\begin{equation}
  \label{eq:deltaU}
  \Delta U_{\text{DVM}}~=~6.412315(15)_\mathrm{stat}(15	)_\mathrm{sys}~\mathrm{V}.
\end{equation}
In the evaluation of the systematic uncertainties associated with this measurement, we considered a $\pm 20$~\% variation of the high-voltage ripple amplitude and a $\pm 50$~\% uncertainty of the synchrotron-radiation correction. The systematic uncertainty of the synchrotron radiation was estimated very conservatively because we did not apply a pixel-wise correction. The assumed $\pm 5$~meV uncertainty on the variation of $U_{\text{pot.dec.}}$ for the different conversion electron lines results in an uncertainty of $\pm 2.5$~$\upmu$V for $\Delta U_{\text{DVM}}$ (equation \ref{u_ret_m}).

In the voltage determination with the DVM, we applied a 0.5~ppm uncertainty on the read value and a 0.2~ppm uncertainty on the full range of the device. These effects yield uncertainties of 8.5~$\upmu$V (11.7~$\upmu$V) for the K-32 (L$_3$-32) voltage reading, and 14.5~$\upmu$V for $\Delta U_{\text{DVM}}$. 

Since the term $q \cdot \Delta U_{\text{pot.dec.}}$ was already absorbed in the fitted data, the scale factor can be determined simply by
dividing equation \ref{eq:delta_ebin} by equation \ref{eq:deltaU}:
\begin{equation}
\label{result_eq}
M_{K35} = 1972.4488(45)_\mathrm{stat}(91)_\mathrm{sys} \approx 1972.449(10).
\end{equation}
This result is in good agreement with the last calibration at PTB (eq. \ref{PTB_value}) within the uncertainties.  With a four-year interval between the two calibrations, the relative deviation amounts to $\Delta M/M = -2(5)$~ppm. This means that the stability of the scale factor is on the ppm-level per year or better, assuming a constant drift. For a typical KATRIN measurement period, which is partitioned in two-month intervals, sub-ppm-stability can be assumed.

The uncertainty of 5~ppm of this new calibration method is dominated by the uncertainty of the difference of the atomic binding energies (relative uncertainty of 4~ppm). This could improve in the next years with more precise spectroscopic measurements or theoretical calculations. The combined relative statistical uncertainty of about 2~ppm can be improved by future measurements with higher statistics during calibration phases at KATRIN. The similarly large uncertainty of the voltage reading could be improved by measuring the two conversion lines in quick succession ($\sim$20~min.) to mitigate the temporal drift effect of the device.

This measurement has also demonstrated that the relative stability of the HV divider is better than 3~ppm in a two-month interval, which significantly surpasses the design specifications.

\section{Conclusion}
In order to achieve the design sensitivity of 0.2~eV/$c^2$ in the neutrino mass measurement, the retarding potential of the main spectrometer of the KATRIN experiment has to be monitored with a precision of 3~ppm over measurement intervals of two months. 
The retarding voltage is measured with two custom-made ultra-precise HV dividers that have to be calibrated regularly. In the past, such calibrations could only be performed at the special metrology laboratories. In this work, a new calibration method is presented, which based on the energy difference of two conversion electron lines produced by the decay of \kr . This method was previously applied with a condensed \kr\ source at the Mainz neutrino mass experiment, but surface and solid-state effects limited the attainable precision. Measurements with gaseous \kr\ at the KATRIN experiment are not affected by these effects, and allow the HV dividers to be calibrated with an uncertainty of $<$5~ppm. We have shown in this paper that such precision is achievable. The measured scale factor of the divider K35 $M_{\text{K35}} = 1972.449(10)$ is in agreement with earlier PTB calibrations.  The results demonstrate the stability and reliability of the K35 HV divider to sub-ppm-levels over the two-month measurement intervals in KATRIN.  This principle of determining the difference of two conversion electron lines with an electrostatic retardation spectrometer, e.g. of MAC-E-filter type, can be applied to other energy lines in other applications.
\begin{acknowledgements}
We acknowledge the support of Helm\-holtz Association (HGF), Ministry for Education and Research BMBF (05A14VK2 and 05A17PM3), Helmholtz Alliance for Astroparticle Physics (HAP), and Helmholtz Young Investigator Group (VH-NG-1055) in Germany; Ministry of Education, Youth and Sport (CANAM-LM2011019, LTT18021), in cooperation with JINR Dubna (3+3 grants) in the Czech Republic; and the Department of Energy through grants DEFG02-97ER41020, DE-FG02-94ER40818, DE-SC0004036, DEFG02-97ER41033, DE-FG02-97ER41041, DE-AC02-05CH11231, and DE-SC0011091 in the United States.
\end{acknowledgements}

\bibliographystyle{spphys}

\bibliography{biblio.bib}

\end{document}